\documentstyle[preprint,aps]{revtex}
\begin{document}
\draft
\title{Local Moments Coupled to a Strongly Correlated Electron Chain}
\author{S. Moukouri, Liang Chen and L. G. Caron}
\address{Centre de recherche en physique du solide and d\'epartement de physique,\\
Universit\'e de Sherbrooke, Sherbrooke, Qu\'ebec, Canada J1K 2R1}
\maketitle

\begin{abstract}
\widetext\leftskip=1cm\rightskip=1cm{\small A 1D model hamiltonian that is
motivated by the recent discovery of the heavy-fermion behavior in the
cuprates of the $Nd_2CuO_4$ type is studied. It consists of $t-J$
interacting conduction electrons coupled to a lattice of localized 
spins through a Kondo
exchange term $J_K$. Exact diagonalization and density matrix
renormalization group methods are used. The latter method is generalized 
to arbitrary densities. 
At half-filling, a spin gap opens
for all $J_K>0$. Away from half-filling$,$ it is shown that, at strong $J_K$%
, the ground state is an unsaturated ferromagnet 
. At weak $J_K$ the system is in a paramagnetic
phase with enhanced RKKY correlations. The importance of self-screening
of the local moments in the
depletion regime is discussed. We argue that these findings transcend
the specifics of the model. }
\end{abstract}



The study of quantum moments embedded in a metallic host is one of the
most active field of condensed matter physics. The main problem is the
understanding of the low-temperature behavior of localized moments,
which are typically rare-earth lanthanide or actinide ions,
interacting with $p$ or $d$
 bands of conduction electrons. It has been argued that
there are two competing effects: the quenching of the local spins through
the Kondo effect and their magnetic ordering due to the RKKY exchange
interaction\cite{varma}. The canonical Hamiltonians of heavy-fermion
compounds are the periodic Anderson lattice model $(PAM)$ 
and the Kondo lattice model $%
(KLM) $. The $PAM$ and the $KLM$ describe a lattice of local
moments coupled to
a wide conduction band. It is widely accepted that this situation prevails
in the so-called heavy-fermion materials\cite{lee}.

The recent discovery of a heavy-fermion phenomenon in the $Ce$-doped
Neodynium cuprate $Nd_{1.8}Ce_{0.2}CuO_4$\cite{brugger}
has led to an increasing interest
in the study of strongly correlated electrons coupled to magnetic moments%
\cite{fulde}. Although these compounds are at least two-dimensional, as a
first step towards their understanding, 
 we report a study of a $1D$
Hamiltonian which retains the basic ingredients of their physics. This
Hamiltonian
describes the interplay between strongly correlated
electrons and a lattice of local moments. The strongly correlated conduction
electron system is described by the $1D$ $t-J$ model. The interaction
between the conduction electrons and the $Nd$ ions is represented by a Kondo
exchange term. Using standard notations, the Hamiltonian is written as
follows:

\begin{eqnarray}
H&=&-t\sum_{i\sigma }(c_{i\sigma }^{+}c_{i+1\sigma }^{}+h.c.)  \nonumber \\
&+&J_H\sum_i({\bf S}_{ic}^{}\cdot {\bf S}_{i+1c}^{}-\frac{n_{ic}n_{i+1c}}4%
)+J_K\sum_i{\bf S}_{ic}^{}\cdot {\bf S}_{if}^{}  \eqnum{1}
\end{eqnarray}

The Hamiltonian (1) is a natural extension of the one-dimensional $KLM$
in the limit of strongly correlated electrons.
We recall that double occupancy is forbidden here. The indices $c$ and $f$
stand for conduction and localized electrons respectively. We set the
hopping parameter $t$ and the Heisenberg coupling $J_H$ equal to $1$. For
those parameters, which are relevant for the cuprates, the ground state of
the $1D$ $t-J$ model is a spin density wave\cite{ogata}. We discuss the
nature of the ground state as the band filling $\rho $ of the conduction
electrons and $J_K$ are varied. 

We use the density matrix renormalization group $(DMRG)$ method of White\cite
{white}. 
The calculations are done using the infinite system method with 3
blocks instead of four as usual. One can immediately realize that 
there is a problem to fix the electron density at a given value at each
iteration. To overcome this, 
we have targeted
the two states whose electron number brackets the desired density
when building the density matrix, 
This algorithm has been successfully checked for the one-dimensional
$t-J$ model\cite{chen}.
 In the present study, we have kept up to $m=140$ states in the
two external blocks. The states were labelled by the electron number and by
the total $z$-component of the spin. The maximum lattice size we have
reached is $75$
. The truncation error defined as $1-p(m)$, where $p(m)$ is the sum of the
eigenvalues of the density matrix of the states kept, was of the order of $%
5\times 10^{-4}$.

At half-filling, the $t-J$ chain reduces to a Heisenberg antiferromagnet. We
have found that the singlet ground state is separated from the first excited
triplet state by a gap for all $J_K>0$. As shown in Fig.1, the $J_K$
dependence of this gap $\Delta $ is very close to that of the Kondo necklace
model $(KNM)$\cite{caron}. It is linear in the strong $J_K$ limit, and, has
probably a non analytic form for small $J_K$. We note that the values of the
gap found here in the weak coupling limit are much larger than that in the $%
KNM$. Thus, the interaction between the conduction electrons, represented
here by the $z$ part of the Heisenberg term after the Jordan-Wigner
transformations, tends to enhance the gap. In order to understand the
character of the gap, we have compared it with the binding energy per site
which is defined as $E_B=\left[ E_G(J_K=0,N)-E_G(J_K,N)\right] /N,$ where $%
E_G(J_K,N)$ is the ground state energy. It can be observed in Fig.$1$ that
in the strong coupling limit, $\Delta $ is larger than $E_B$: the
excitations of the system are obtained by breaking up on-site singlets. In
the weak coupling regime, $E_B$ becomes greater than $\Delta $, this means
that the system can be excited without the destruction of local singlets.
The gap results from a collective effect. In the $KNM$\cite{caron}, as the
coupling is decreased, the $RKKY$ correlations become larger in magnitude.
The situation is also similar here. Fig.$2a$ displays the magnetic structure
factors, $S_{f,c}(k)=\frac 4N\sum_{l,m}<S_{f,c,l}^zS_{f,c,m}^z>\exp
[i(l-m)k] $, of localized and conduction electrons respectively. These
structure factors have a maximum at 2k$_{Fc}$= $\pi $ for $J_K=0.2$. When
the coupling is increased, the cusp flattens out, indicating the vanishing
of the $RKKY$ correlations. We emphasize that even in the $RKKY$ regime,
there is no long-range order or even a power-law decay of the correlations
because the tendency to magnetic order is thwarted by the gap induced by the
Kondo mechanism. Finally, we note that our results are in agreement with
those of Igarashi et al.\cite{fulde2} who also studied the effect of lattice
of local moments on a Heisenberg chain using exact an diagonalization method.

Away from half-filling, the situation is less clear because the spin
configuration of the ground state is not known. Actually, this information
is important in a numerical calculation since it allows a significant
reduction of the Hilbert space under study. There are however two limits
where one can gain some insight. The first one is the situation with only
one conduction electron, which is identical to the $KLM$ with one electron.
The theorem of Sigrist et al.\cite{sigrist} applies: the ground state is an
unsaturated ferromagnet having a total spin $S_T=(N-1)/2$. The second one is
the limit of infinite coupling, in which all the conduction electrons form
on-site singlets with the localized spins, so that $N-N_c$ spins remain
free. The ground-state has $2^{(N-N_c)}$-fold spin degeneracy. In this
limit, the degrees of freedom consist in local singlets and unpaired $f$ (up
or down) spins. The spin degeneracy can be partially lifted by the
introduction of the hopping term. The Heisenberg term has no role because
there are no free conduction electrons. This situation is identical to the
strong coupling limit of the $KLM$, where Sigrist et al.\cite{sigrist2} have
demonstrated that the ground state is a ferromagnet with $S_T=(N-N_c)/2$ for
all $\rho \neq 1$. Their result also applies in the present case. In order
to clarify the behavior of the system in the whole range of parameters, we
have performed an exact diagonalization for lattice sizes of $4,5,6$ sites with
open boundary conditions. We have found that there is a critical value of
the Kondo coupling $J_c$, which depends on the band filling. When $J_K\geq
J_c$, the ground-state is always a ferromagnet at all $\rho \neq 1$. Below
this transition point, the spin configuration of the ground-state evolves
from the maximum value of the spin to the minimum. There is a transition
region where the spin of the ground state can have an intermediate value.
But at smaller $J_K$, finite size effects become important and trends are
less clear. We display in Fig. 3 the energy difference $\delta E$ between
the triplet and the singlet lowest energy states for $N=4,5,6$ sites with $%
N_c=2,3,4$ electrons respectively. For the 4 sites case, $\delta E$ is always
positive below the transition point, while for $5$ and $6$ sites its sign is
not constant reflecting the finite size effects. We also present the
interesting case of $2$ electrons in $6$ sites. It can help us understand
the evolution from the maximum spin ground state $S_T=2$ to the minimum $%
S_T=0$. At strong coupling, the ground state has $S_T=2$, there is a narrow
region near the transition point where the spin configuration of the ground
state is $S_T=1$. Then it becomes $S_T=0$ at smaller couplings. One can also
observe that when the density is reduced, the value of $J_c$ decreases.
Having in mind these results, we use the $DMRG$ to compare the lowest energy
states with $S_T=0$ and $(N-N_c)/2$. Since in the $DMRG$, the states are
labelled by only the $z$-component of the total spin, at each value of the
coupling, we have calculated the lowest energy of the states with $%
S_T^z=0,\pm 1$ and $\pm (N-Nc)/2$, such that the last number is an integer.
When these energies are all different, we have found that the lowest value
is that of the state with $S_T^z=0$. It is thus reasonable to think that
the ground-state is a singlet. This occurs when $J_K<J_c$. But when $J_K>J_c$
all the states have the same energy ( in the order of $10^{-3}$) and we
conclude that the ferromagnetic state is preferred. These findings
corroborate the exact diagonalization calculations, namely that there is a
cross-over from a singlet ground state to a ferromagnet at the densities
that we have investigated. We caution that in the finite size study there is
a transition region near $J_c$ where the ground state can have an
intermediate value of the total spin. A detailed study of this region is
left for the future.

Let us now reexamine the magnetic structure factors of the $f$ electrons
shown in Fig. $2$. In the strong coupling limit, $J_K=2$, starting from
half-filling (Fig. $2a$) where each $f$ electron is screened by a conduction
electron, the system is in non-magnetic state. The reduction of the density
is equivalent to the insertion of $N-N_c$ free spins within $N_c$ singlets.
The problem is analogous to the insertion of holes in the strong $U$ limit
of the Hubbard model. At $\rho =0.9$ (Fig. $2b$), we found that the
ground-state has $S_T=(N-N_c)/2$, thus, contrary to the strong $U$ limit of
the $1D$ Hubbard model, the ground state is an unsaturated ferromagnet; this
extends the results of exact diagonalization to larger systems. Although the
system is in the ferromagnetic state, the cusp at $k=0$ is very small, we
believe that this is due to the fact that the density of the singlets is
still much larger than that of the unpaired $f$ electrons, so that
ferromagnetic correlations are hidden since the effective coupling between
the unpaired $f$ electrons is small\cite{sigrist2}. As one decreases the
band filling (Fig. $2c,2d$), the maximum at $k=0$ steeply rises. The lower
is the density, the higher is the maximum, since the number of the ordered
spins is greater. At intermediate couplings, the $FM$ state is destroyed for 
$\rho =0.9$ around $J_c=1.2$ and $\rho =0.7$, around $J_c=1$. One can
clearly see that there is no dominant feature in $S_f(k)$ in these two cases
when $J_K=1$. But the $FM$ state is still stable at $\rho =0.5$ as indicated
by the maximum at $k=0$, the transition occuring only at $J_c=0.8$. Thus,
the $FM-PM$ phase boundary is shifted towards smaller $J_K$ when the band
filling is decreased. This finding is consistent with the existence of
ferromagtism for all $J_K>0$ of the case with one electron. In the small
coupling regime, the ground state is always a singlet. An incommensurate
peak appears in $S_f(k)$ at $2k_{Fc}=\pi \rho $ at the three densities
studied, showing that the $RKKY$ correlations are dominant.

In the weak $J_K$ region we see the ground state is still a singlet despite $%
Nc<N$. 
 Nozi\`eres \cite{nozieres} has argued that at low
temperatures, there will not be enough conduction electrons to screen out
all the localized spins. This incomplete Kondo screening could lead to an
ineffective $RKKY$ interaction. We see that in our case the nature of the
screening process seems to be very different from that of the one-impurity
case since a spin-compensated state exists even though $N_c<N$. A scenario
discussed in\cite{shiba} , to avoid the Nozi\`eres exhaustion problem, was
that a significant part of screening is done by the $f$ electrons
themselves, through the action of intersite $RKKY$ correlations. Following
Blankenbecler et al.\cite{scalapino} the condition for the quenching of a
spin located at the origin gives rise to the compensation sum rule:

\begin{equation}
<S_{f,0}^{z2}>=-\sum_{l\neq 0}<S_{f,l}^zS_{f,0}^z>-\sum_l<S_{c,l}^zS_{f,0}^z>
\eqnum{2}
\end{equation}

We have verified this relation for $J_K=0.5$ where a Kondo state is expected
since the ground state is a singlet and the
magnetic correlations are short ranged 
. The left hand side of $(2)$ is
always equal to $0.25$. The first term of the rigth-hand side of $(2)$,
which gives the contribution of the $f$ electrons to the screening slightly
increases when the density is reduced. It takes the values $0.08,0.08,0.11$
and $0.12$ for $\rho =1,0.9,0.7$ and $0.5$ respectively when the $10$ first
neighbors are included. At the same time, the contribution of the conduction
electrons which is measured by the second term decreases. It is equal
respectively to $0.12,0.11,0.10$ and $0.08$. Even at half-filling where the
number of conduction electrons is equal to that of the local moments
, the
``self-screening'' is non negligible. However, when $J_K$ is increased,
strong on-site singlets are formed and this screening mechanism is no longer
possible because $RKKY$ correlations between $f$ electrons are suppressed.
The major contribution to the compensation sum rule comes from the on-site
electron-spin correlation $-<S_{c,0}^zS_{f,0}^z>$ which is displayed in Fig. 
$4$. This quantity is found to be very close to $0.25\rho $ in the strong
coupling regime. Therefore, in the large $J_K$ limit $(2)$ is satisfied only
at half-filling, since the $f$ electrons are $FM$ ordered when $\rho \neq 1$%
. Thus the depletion effects occur because the formation of a coherent
singlet ground state necessitates the action of the antiferromagnetic
RKKY correlations.

So far, we have mainly discussed the action of the electrons on the
 lattice of local moments. 
We now wish to consider the inverse problem. For the
conduction electrons, we have found that $S_c(k)$ is flat at strong and
intermediate couplings for $\rho =0.9$ (Fig$.2b$). We note that for $\rho
=0.5$ and $\rho =0.7$, a small peak arises at $k=0$. This is a 
consequence of the $f$-spin induced ferromagnetic correlations of the
conduction electrons (Fig. $2c,2d$).
Dominant structures appear in the structure factor $S_c(k)$ at $2k_{F_c}$
only for small $J_K$.

In Summary, we have studied a model Hamiltonian which describes the physics
of magnetic moments coupled to strongly correlated electrons in
1D. We have found that at half-filling, the system is a spin gaped insulator
for all non zero values of the Kondo coupling. When the system is doped, a
FM state is stable at strong coupling. We have argued that it is a
consequence of the depletion effects. At weak coupling, the ground state is
a collective singlet, we have shown that a significant part of the screening is made by
the local spins themselves through the RKKY interaction. Finally, we have    
reached the same kind of conclusion in a recent study of the 
ground-state properties of the one-dimensional
$KLM$\cite{moukouri}. It seems that the Hamiltonian $(1)$ and the $KLM$
belong to the same universality class. Indeed, 
in the $KLM$, there is no explicit
interaction between the conduction electrons. However, it is well-known
that the Kondo exchange term induces an effective interaction between
these electrons. It is thus likely that the Hamiltonian (1) and the $KLM$
are not fundamentally different.

We wish to thank A.-M. S. Tremblay and C. Bourbonnais for useful
discussions. This work was supported by a grant from the Natural Sciences
and Engineering research Council (NSERC) of Canada and the Fonds pour la
formation de Chercheurs et l'Aide \`a la Recherche (FCAR) of the Qu\'ebec
government.

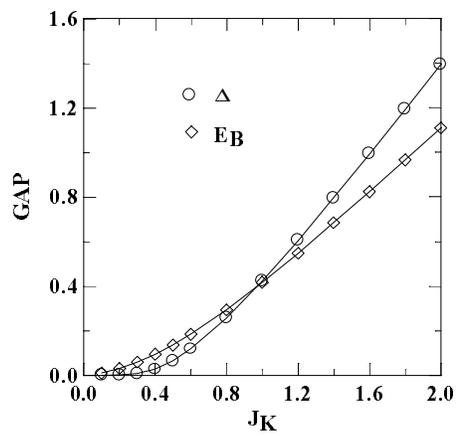
\begin{figure}[tbp]
\caption{ The singlet triplet gap $\Delta $ (circles) and the binding energy 
$E_B$ (diamonds) per site versus the Kondo coupling.}
\end{figure}
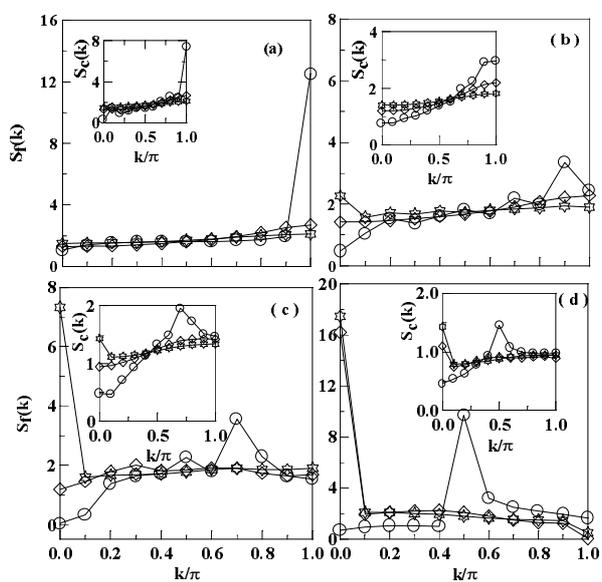
\begin{figure}[tbp]
\caption{ The magnetic structure factor of the localized spins and of the
conduction electrons (inset) for $(a)$ $\rho =1,(b)$ $\rho =0.9,$ $(c)$ $%
\rho =0.7,$ $(d)$ $\rho =0.5$ at $J_K=0.2$ (circles), $J_K=1$ (diamonds) and 
$J_K=2$ (stars).}
\end{figure}
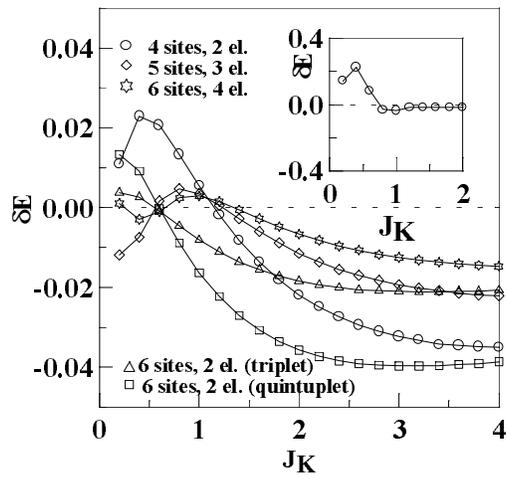
\begin{figure}[tbp]
\caption{ Energy difference $\delta E$ between the lowest states with $S_T=1$
(or $2$) and $S_T=0$ for systems of $4,5$ and $6$ sites with $2,3$ and $4$
(or $2$) electrons respectively. In the inset, $\delta E$ between the lowest
states with $S_T^z=0$ and $S_T^z=(N-N_c)/2$ for $\rho =1/2$.}
\end{figure}
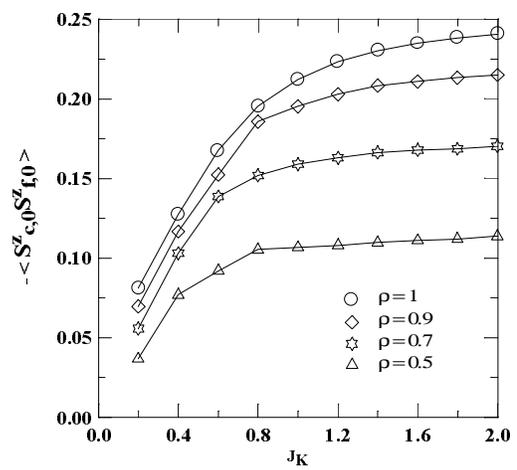
\begin{figure}[tbp]
\caption{The on-site electron-spin correlation versus the Kondo coupling at $%
\rho =1$ (circles), $\rho =0.9$ (diamonds), $\rho =0.7$ (stars) and $\rho
=0.5$ (triangles).}
\end{figure}

\end{document}